\documentclass[prd,aps,showpacs,preprintnumbers,amsmath, amssymb]{revtex4-2}

\usepackage{graphics,epsfig,hyperref}
\usepackage{graphicx}
\usepackage{dcolumn}
\usepackage{bm}
\usepackage{epstopdf}
\usepackage{color}
\usepackage{xcolor}
\usepackage{subfigure}
\usepackage{diagbox} 
\usepackage{adjustbox}
\usepackage{multirow}
\usepackage{amsmath,amssymb,amsfonts}

\usepackage[normalem]{ulem}
\usepackage{xcolor}

\usepackage{comment}

\begin{document}
	\title{\boldmath 
	Bayesian inference for tidal heating with extreme mass ratio inspirals}
	
	\author{ Zhong-Wu Xia$^{1}$\footnote{zwxia@hunnu.edu.cn}, Sheng Long$^{2}$\footnote{shenglong@ucas.ac.cn}, Qiyuan Pan$^{1,3}$\footnote{panqiyuan@hunnu.edu.cn}, Jiliang Jing$^{1,3}$\footnote{jljing@hunnu.edu.cn} and Wei-Liang Qian$^{4,5,3}$\footnote{wlqian@usp.br}}
	\affiliation{$^1$Department of Physics, Institute of Interdisciplinary Studies, Key Laboratory of Low Dimensional Quantum Structures and Quantum Control of Ministry of Education, Synergetic Innovation Center for Quantum Effects and Applications, and Hunan Research Center of the Basic Discipline for Quantum Effects and Quantum Technologies, Hunan Normal University,  Changsha, Hunan 410081, People's Republic of China} 
	\affiliation{$^2$School of Fundamental Physics and Mathematical Sciences, Hangzhou Institute for Advanced Study, UCAS, Hangzhou 310024, People's Republic of China}
	\affiliation{$^{3}$Center for Gravitation and Cosmology, College of Physical Science and Technology, Yangzhou University, Yangzhou 225009, People's Republic of China}
	\affiliation{$^{4}$Escola de Engenharia de Lorena, Universidade de São Paulo, 12602$-$810, Lorena, SP, Brazil}
	\affiliation{$^{5}$Faculdade de Engenharia de Guaratinguet\'a, Universidade Estadual Paulista, 12516-410, Guaratinguet\'a, SP, Brazil}

	\begin{abstract}
	\baselineskip=0.6 cm
	\begin{center}
		{\bf Abstract}
	\end{center}

Extreme mass ratio inspirals (EMRIs) provide unique probes of near-horizon dissipation through the tidal heating. We present a full Bayesian analysis of tidal heating in equatorial eccentric EMRIs by performing injection–recovery studies and inferring posterior constraints on the reflectivity parameter $|\mathcal{R}|^2$ while sampling in the full EMRI parameter space.  We find that in the strong-field regime the posterior uncertainties are smaller, indicating a stronger constraining capability on the tidal heating. Using two-year signals with an optimal signal-to-noise ratio (SNR) of $\rho=50$, EMRIs can  put bounds on  $|\mathcal{R}|^2$ at the level of $10^{-3}$--$ 10^{-4}$ for a rapidly spinning central object. Moreover, we  show that neglecting the tidal heating can induce clear systematic biases in the intrinsic parameters of the EMRI system. These results establish EMRIs as promising precision probes for detecting and constraining black hole event horizons.

    \end{abstract}

\maketitle
\flushbottom

\section{Introduction}
\label{sec:intro}

Gravitational wave (GW) observations over the past decade, primarily from ground-based interferometers, have opened a new window on compact-object binaries and enabled precision tests of gravity in the strong-field regime~\cite{LIGOScientific:2025obp,LIGOScientific:2025rid}. Planned space-based GW detectors, including LISA~\cite{LISA:2024hlh}, Taiji~\cite{Ruan:2018tsw} and TianQin~\cite{TianQin:2015yph}, will access the millihertz band and observe sources over months to years, allowing the SNR to build up to high values.
An EMRI system, in which a stellar-mass compact object slowly spirals into a massive black hole (MBH), is one of the most  important  sources  of space-based  GW detectors, offering a direct probe of the spacetime geometry around MBHs.
Unlike comparable-mass binaries observed by ground-based detectors, which spend only a few tens of cycles in the strong-field regime, EMRIs remain in LISA's band for years and can accumulate about $10^{6}$ orbital cycles  in the strong-field regime~\cite{Barack:2004hh,AmaroSeoane:2018eoq,LISAConsortiumWaveformWorkingGroup:2023arg}.
As a result, even small modifications to the energy and angular momentum fluxes can lead to observable dephasing in the GW waveform~\cite{Hartle:1973zz,Hughes:2001jr,Datta:2020oph}.
Therefore, EMRIs will offer unprecedented insights into testing general relativity (GR): they provide a unique opportunity to detect deviations from Kerr black hole predictions~\cite{Li:2022pcy,Liu:2023onj,Zhang:2024csc}, including constraints on modifications to the gravity~\cite{Cardenas-Avendano:2024mqp,Qiao:2024gfb,Kumar:2025njz,Fu:2024cfk,Xia:2025yzg,Zhao:2025sck,Yang:2024cnd,Yang:2025esa,Chen:2026kbn,Liu:2025swi,Hua:2026kvw,Guo:2025scs,Dong:2026xab,Guo:2024bqe,Zhang:2026lrd} and potential beyond-vacuum environments, such as the dark matter~\cite{Duque:2023seg,Li:2021pxf,Dyson:2025dlj,Li:2025ffh,Mitra:2025tag,Xiong:2025zqp,Feng:2025fkc,Meng:2025glf,Destounis:2025tjn}, accretion disks~\cite{Pan:2021oob,Speri:2022upm,Li:2025zgo,Zi:2026zpw}, and additional fields~\cite{Barsanti:2022vvl,Liang:2022gdk,Zi:2024lmt,Zi:2025onl}. Moreover, they enable a detailed characterization of the secondary’s structure~\cite{Lestingi:2023ovn,Cui:2025bgu,Guo:2023mhq}.

In addition to the applications mentioned above, EMRIs provide a precise probe of near-horizon dissipation, namely the transfer of energy and angular momentum from the orbit into the central object through GWs absorption determined by the horizon boundary condition, a phenomenon commonly referred to as the tidal heating~\cite{Munna:2023vds}. In GR, the event horizon acts as a one-way membrane and can absorb the radiation near the horizon~\cite{Poisson:2004cw,Chatziioannou:2012gq,Chatziioannou:2016kem}. In a binary system, this absorbed flux can be equivalently described as a tidal deformation of the horizon that backreacts on the orbital dynamics. The effect becomes particularly significant for rapidly spinning black holes and is most pronounced in the late stages of the inspiral. Moreover, the rotation introduces an additional channel: a spinning black hole can amplify low-frequency radiation through the superradiance~\cite{Brito:2015oca,Yang:2022uze}, making the horizon behave as a dissipative system analogous to a Newtonian viscous fluid~\cite{Poisson:2009di}. In the superradiant regime, the energy and angular momentum are secularly extracted from the BH spin and transferred to the orbit; outside this regime, the horizon term instead contributes additional dissipation. In either case, the cumulative effect can lead to substantial dephasings in the LISA band~\cite{Datta:2024vll}.

By contrast, quantum gravity corrections to the event horizon~\cite{Maselli:2017cmm,Bianchi:2020miz,Agullo:2020hxe}, or  phase transition~\cite{Mazur:2004fk}, can both give rise to exotic compact objects (ECOs) that closely mimic BHs: their exterior geometry is indistinguishable from the Kerr spacetime, but the boundary conditions at the horizon  differ from  the Kerr BH case ~\cite{Cardoso:2019rvt}.
A convenient, model-agnostic parameterization is to  introduce a near-surface reflectivity $\mathcal R$ that rescales the  horizon contribution, i.e., $\dot E_{\rm tot}=\dot E_\infty+(1-|\mathcal R|^2)\dot E_{\rm H}$ (and similarly for $\dot L$), with $\mathcal R=0$ recovering a perfectly absorbing Kerr BH and $|\mathcal R|=1$ corresponding to an idealized perfectly reflecting surface~\cite{Datta:2024vll}.
In this sense, altering the near-horizon boundary condition  modifies the horizon flux and the associated secular tidal heating, even when the exterior metric is nearly Kerr, thereby leading to the dephasing compared with the Kerr case. Previous works have investigated the detectability of tidal heating by using  EMRIs.
In Ref.~\cite{Zi:2024itp}, Zi \textit{et al.} combined post-Newtonian (PN) fluxes with analytic augmented kludge waveforms. This allows them to study the tidal heating for eccentric and inclined orbits and to perform a Fisher matrix analysis. However, the kludge waveforms are less accurate in the strong-field regime where the tidal heating is most important~\cite{Datta:2024vll}.
In Refs.~\cite{datta2020tidal,Datta:2024vll}, Datta \textit{et al.} considered fully relativistic fluxes and waveform models for equatorial circular and eccentric orbits, and quantified the sensitivity through a mismatch analysis. In addition, it is worth noting that in Ref.~\cite{zi2024detecting}, Zi \textit{et al.} modelled the boundary condition at the would-be horizon as a superposition of ingoing and outgoing waves to study EMRI signals in ECO spacetimes. Together, these approaches provide valuable guidance and useful computational benchmarks.
However, these earlier studies did not place rigorous posterior limits on the detectability of tidal heating. To fully achieve the scientific goals of space-based GW detectors for EMRIs, it is necessary to infer the population distribution of their physical parameters from the data.

In this work we are primarily interested in assessing the detectability of tidal heating in EMRI observations and in placing quantitative constraints on its magnitude. 
Previous work has preliminarily assessed  LISA’s sensitivity to tidal heating effects by using the mismatch  and Fisher  information matrix. Here we instead adopt a fully Bayesian analysis, producing samples to infer the posterior distribution of tidal heating  parameter alongside the standard source parameters.
In our calculation, we use the Markov Chain Monte Carlo (MCMC) method proposed for EMRIs~\cite{Wang:2019ryf}. To model the tidal heating accurately, we adopt \textbf{FastEMRIWaveforms}~\cite{chapman_bird_2025_15630565}, which enables a fully Bayesian analysis  in equatorial eccentric EMRIs by using fully relativistic adiabatic waveforms.
For the first time, we infer the posterior distribution of tidal heating  through the   sampling with a precise zero-order post-adiabatic (PA) waveform model and full LISA response, and assess the detectability of tidal heating in EMRI observations and potential bias from neglecting the tidal heating.

The work is organized as follows. In Sec.~\ref{sec:2}, we describe the adiabatic inspiral and waveform construction, including how the reflectivity enters the horizon flux and how we generate the detector response. In Sec.~\ref{sec:3}, we present the Bayesian inference framework, specify the likelihood and prior choices, and show the MCMC sampling details. We then report the resulting posterior constraints, parameter correlations and the dependence of measurability across representative orbital configurations. Finally, in Sec.~\ref{sec:4}, we summarize our main conclusions and discuss implications for testing the horizon dissipation with future space-based GW observations.
Throughout this paper, we use geometric units with $G = c = 1$.

\section{Adiabatic Inspiral and Waveform}
\label{sec:2}

Over the past two decades, the EMRI waveform modelling for LISA has progressed from fast but approximate “kludge'' templates~\cite{Barack:2003fp,Babak:2006uv,Chua:2017ujo} to adiabatic models rooted in Kerr geodesics and BH perturbation theory. Kludge waveforms are convenient for survey studies, but their semi-relativistic treatment of the orbit leads to uncontrolled phase errors over multi-year inspirals, particularly in the near-horizon region where the tidal field is strongest. 
In principle, an EMRI waveform model accurate to first PA order is required for the precision inference of parameters. However, owing to current technical challenges, fully relativistic adiabatic waveforms in Kerr spacetime are at present only available up to the 0PA order~\cite{Mathews:2025txc,Zhang:2025eqz,Wei:2025lva}.
In this work, we therefore adopt the fully relativistic 0PA waveform model of Ref.~\cite{chapman2025fast} for computationally efficient large-scale analyses.

Here we  briefly review how to include tidal heating corrections in the adiabatic waveform.
In this work we consider Kerr-like  ECOs whose exterior geometry is well described by the Kerr metric, while the event horizon is replaced by an effective, partially reflecting surface.
In the adiabatic EMRI regime, this motivates using standard Kerr geodesic dynamics and Teukolsky-based method for the dissipative evolution, and encoding the ECO imprint only through a modification of the near-horizon flux.
Since the mass ratio is of order $10^{-5}$, the small BH can therefore naturally be treated as a perturbation of the background spacetime, with the metric expanded as
$
g_{\mu\nu}=g^{(0)}_{\mu\nu}+h^{(1)}_{\mu\nu}+\mathcal O(h^{(2)}_{\mu\nu}),
$
where $g^{(0)}_{\mu\nu}$ is the  background metric and $h^{(1)}_{\mu\nu}$ is a first-order perturbation. Consequently, at leading order in the mass ratio, the system is accurately described by the linearized Einstein equations
\begin{equation}
	G_{\mu\nu}^{(1)}=8\pi\kappa T_{\mu\nu}^{(1)}.
\end{equation}
For a Kerr-like ECO, which is a Ricci-flat spacetime of Petrov type D, one can  first recast the linearized Einstein equations in the Newman–Penrose formalism~\cite{Newman:1961qr,Chandrasekhar:1985kt} or Geroch-Held-Penrose formalism~\cite{Geroch:1973am, Pound:2021qin,price2007developments,Aksteiner:2014zyp}, leading to the Teukolsky equation. The GWs observed by an asymptotic observer at infinity are described by the Weyl scalar $\Psi_4^{(1)}$.
The separation of variables in the Boyer-Lidquist coordinate
\begin{equation}
		(r - i a \cos\theta)^{4} \Psi_{4}^{(1)}
	= \sum_{\ell m} \int d\omega\,
	e^{- i \omega t + i m \varphi}\,
	{}_{-2}S_{\ell m}(\theta)\, {}_{-2}R_{\ell m \omega}(r) \, 
\end{equation}
 reduces the Teukolsky equation to the radial equation and the angular equation for spin-weighted spheroidal harmonics~\cite{teukolsky1973perturbations,mino1997chapter}
\begin{align}
		\Delta^{2}\,\frac{d}{dr}\!\left(\frac{1}{\Delta}\,\frac{d}{dr}\right){}_{-2}R_{\ell m \omega}
	-\left[-\,\frac{K^{2} + 4 i (r - M) K}{\Delta} \;+\; 8 i \omega r \;+\; \lambda \right]\,{}_{-2}R_{\ell m \omega}
	= T^{(1)}_{\ell m \omega},\label{Teu}	\\
	\left[
	\frac{1}{\sin\theta}\frac{d}{d\theta}
	\left( \sin\theta \frac{d}{d\theta} \right)
	- a^{2}\omega^{2}\sin^{2}\theta
	- \frac{(m - 2\cos\theta)^{2}}{\sin^{2}\theta}
	+ 4a\omega\cos\theta
	- 2 + 2ma\omega + \lambda
	\right]\; {}_{-2}S_{\ell m} = 0,\label{swh}
\end{align}
with $\Delta = r^2-2Mr+a^2$ and $K=(r^2+a^2)\omega-ma$.
$T^{(1)}_{\ell m \omega}$ is the source term in the Einstein field equations, describing the motion of a point particle in the ECO spacetime, which has been given by Ref.~\cite{Drasco:2005kz}. 
The solution of the angular equation is the spin-weighted spheroidal harmonics with the spin weight $-2$, which admits straightforward numerical solutions~\cite{BHPToolkit,Hatsuda:2020sbn}. 
For the  homogeneous part  of the radial  equation \eqref{Teu}, a progression of methods has been developed~\cite{mano1996analytic,fujita2004new,hughes2000evolution,glampedakis2002zoom,Harms:2013ib,He:2024drw,Lo:2023fvv,Jiang:2025mna,Yin:2025kls}.
A good practice is to choose different methods for different modes depending on their domains of applicability, and thereby obtain a complete set of fluxes and waveforms~\cite{Nasipak:2022xjh,Nasipak:2023kuf,Nasipak:2025tby}.
To describe the energy-momentum tensor $T^{(1)}_{\ell m \omega}$, it is convenient to introduce the Keplerian–like orbit for the bound  orbits~\cite{Babak:2006uv}
\begin{equation}
	r(\psi) = \frac{p}{1 + e\cos\psi},
\end{equation}
where $e$ is the eccentricity and $p$ is the semi-latus rectum, and $\psi$ parameterizes the radial phase of the orbit. From which the inhomogeneous solutions are then assembled rapidly and accurately by the Green's function method. Finally, the GW corresponding to an equatorial orbit, as detected at infinity, is given by
\begin{equation}\label{eq:psi4_infty}
	h_+-ih_\times
	= \frac{1}{r}\sum_{lmn} Z_{lmn}^{\infty}
	{}_{-2}S_{lm}\big(\theta; a\omega_{mn}\big)
	e^{i\big(m\phi-\omega_{mn}(t-t_i)\big)},
\end{equation}
where $Z^\infty_{lmn}$ denotes the GW amplitude at infinity and $\omega_{mn} = m\Omega_\phi + n\Omega_r$ is the corresponding GW frequency. For space-based GW detectors like LISA, the polarization modes of the GW are mapped into the detector response
\begin{equation}
	h_{\text{I,II}}=\frac{\sqrt 3}{2}(F^+_{\text{I,II}}h_++F^\times_{\text{I,II}}h_\times).
\end{equation}
The definition of $F_{\text{I,II}}$ can be found in~\cite{Cutler:1994ys}. 

EMRIs emit GWs during their evolution, carrying away the energy and angular momentum from the system. Meanwhile, motivated by earlier studies, which indicate that the tidal heating becomes non-negligible as the inspiral approaches the strong field near the separatrix regime, we treat the  reflectivity $\mathcal{R}$ as an additional intrinsic parameter that enters the horizon dissipation of the EMRI dynamics.
In the adiabatic approximation, the total time-averaged rates of change of the gravitational energy and angular momentum  can be written as~\cite{Teukolsky:1974yv,datta2024tidal}
\begin{align}
	\frac{dE}{dt}=\left(\frac{dE}{dt}\right)^{\infty}+(1-|\mathcal R|^2)\left(\frac{dE}{dt}\right)^{H}
	&= \sum_{lmn} \frac{|Z^{\infty}_{lmn}|^{2}}{4\pi \omega_{mn}^{2}}+(1-|\mathcal R|^2)\sum_{lmn} \alpha_{lmn}\frac{|Z^{H}_{lmn}|^{2}}{4\pi \omega_{mn}^{2}},\\
	\frac{dL}{dt}=\left(\frac{dL}{dt}\right)^{\infty}+(1-|\mathcal R|^2)\left(\frac{dL}{dt}\right)^{H}
	&= \sum_{lmn} \frac{m |Z^{\infty}_{lmn}|^{2}}{4\pi \omega_{mn}^{3}}+(1-|\mathcal R|^2) \sum_{lmn} \alpha_{lmn}\frac{m |Z^{H}_{lmn}|^{2}}{4\pi \omega_{mn}^{3}}. 
\end{align}
The definition of $\alpha_{lmn}$ can be found in Ref.~\cite{Fujita:2009us}. Here, the partial absorption is implemented by rescaling the horizon  contribution
to the dissipative fluxes by a factor $(1-|\mathcal R|^2)$, such that $\mathcal R=0$ reproduces the
BH limit while $|\mathcal R|=1$ suppresses the horizon dissipation. Since we only consider equatorial orbits, the flux of the Carter constant vanishes, and the inspiral is fully characterized by the energy and angular momentum fluxes.

\begin{figure}[tbp]
	\centering
	\begin{minipage}[b]{0.48\textwidth}
		\centering
		\includegraphics[width=\textwidth]{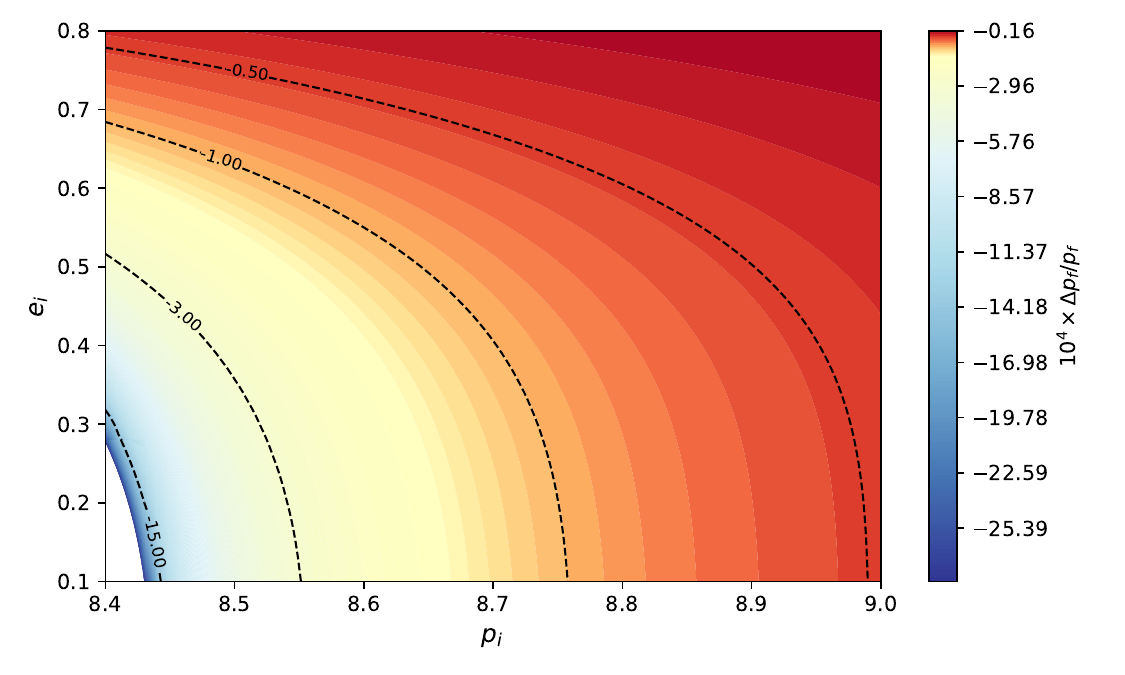}
	\end{minipage}
	\hfill
	\begin{minipage}[b]{0.48\textwidth}
		\centering
		\includegraphics[width=\textwidth]{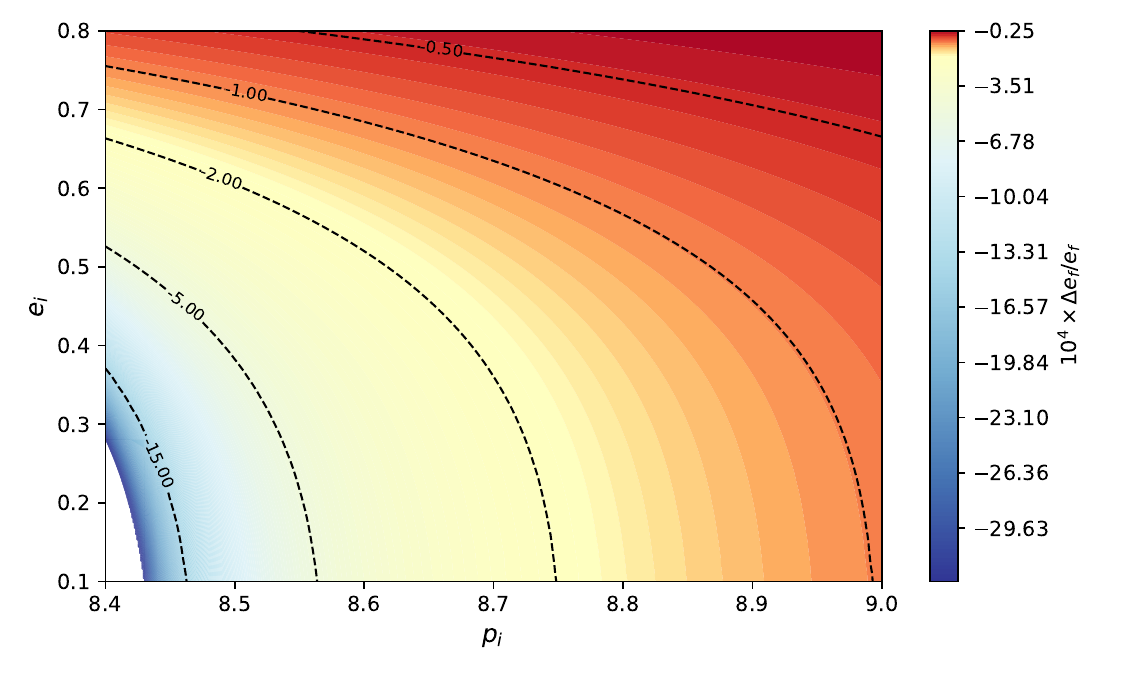}
	\end{minipage}
	\vspace{0.5cm} 
	\begin{minipage}[b]{0.48\textwidth}
		\centering
		\includegraphics[width=\textwidth]{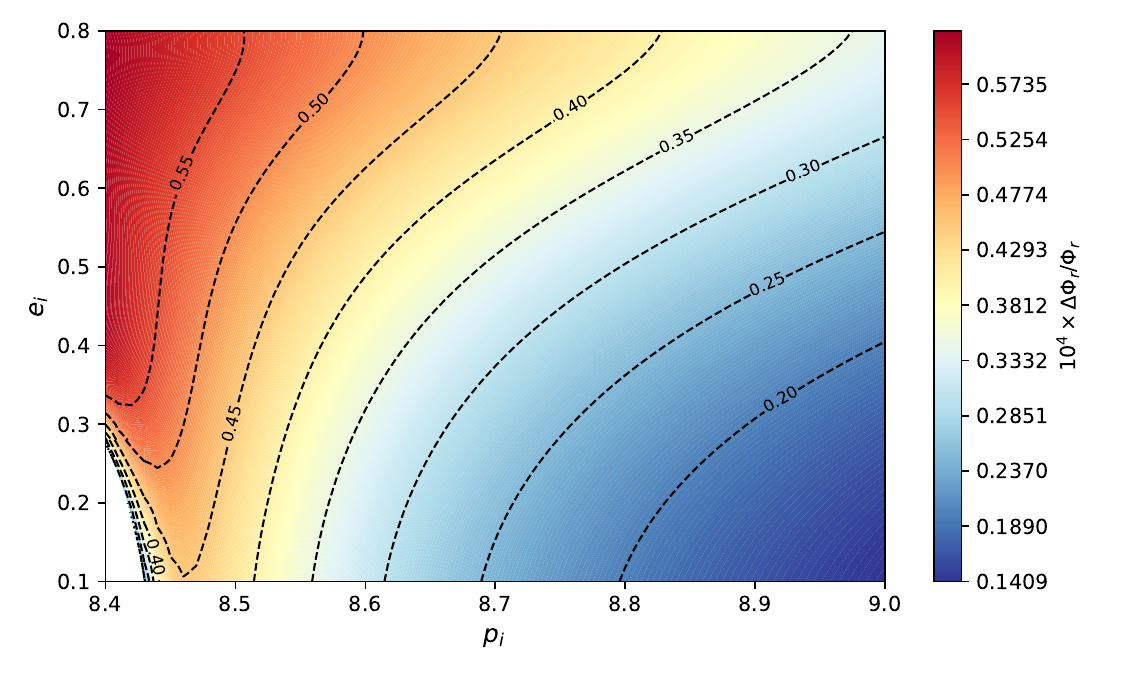}
	\end{minipage}
	\hfill
	\begin{minipage}[b]{0.48\textwidth}
		\centering
		\includegraphics[width=\textwidth]{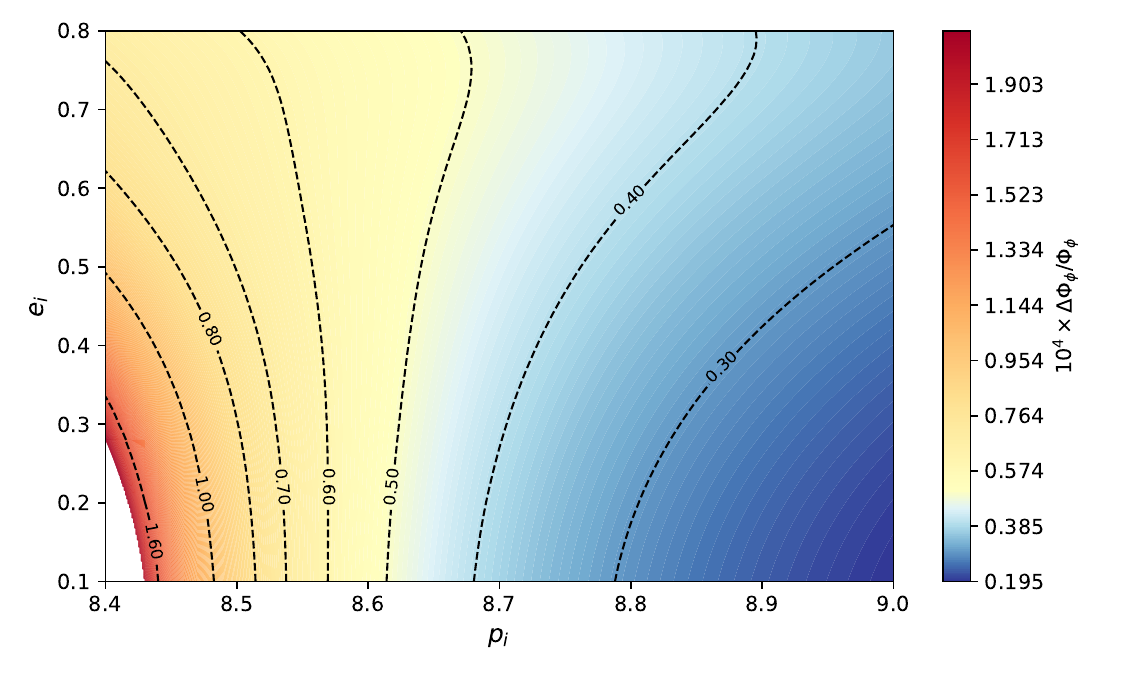}
	\end{minipage}

\caption{The change to the inspiral trajectory arising from the tidal heating. We fix the evolution time to $T=2~\mathrm{yr}$ and compare two runs with the reflectivity $|\mathcal{R}|^2=0$ and $|\mathcal{R}|^2=0.01$ over the initial-condition grid $(p_i,e_i)$ for $(M,\mu,a)=(10^6 M_\odot,\,10\,M_\odot,0.95M)$.
	The four panels are arranged as follows: \emph{upper left} and \emph{upper right} show the fractional change in the final semi-latus rectum $10^4\times(p_f(0.01)-p_f(0))/p_f(0)$ and the final eccentricity $10^4\times(e_f(0.01)-e_f(0))/e_f(0)$, and \emph{lower left} and \emph{lower right} show the relative change in the accumulated radial phase  $10^4\times(\Phi_r(0.01)-\Phi_r(0))/\Phi_r(0)$ and the azimuthal phase $10^4\times(\Phi_\phi(0.01)-\Phi_\phi(0))/\Phi_\phi(0)$, respectively. 
}

	\label{fig:orbit}

\end{figure}

The reflectivity $|\mathcal R|^2$ is an intrinsic parameter because it  modifies the horizon  fluxes and therefore the secular phasing.
As a consequence, even very small values of $|\mathcal R|^2$ can generate a coherent, cumulative phase shift as the inspiral approaches the strong field near the separatrix regime, where the horizon flux contribution is strongest.
This intrinsic and long-time imprint is precisely why  the mismatch analysis in early work finds that the sensitivity to reflectivities can reach
$|\mathcal R|^2\sim \mathcal O(10^{-5})$. Since our main goal is to constrain $\mathcal R$ with a two-year Bayesian  analysis, we fix the observation time to $T=2\,{\mathrm yr}$ and consider a small fiducial deviation in the reflectivity,
then examine how this perturbation propagates into both the final orbital parameters $(p_f,e_f)$ and the accumulated
phases $(\Phi_r$, $\Phi_\phi)$ over the full two-year evolution in Fig.~\ref{fig:orbit}.
In each panel, the white region in the lower-left corner marks initial conditions whose plunge time is shorter than $2\,{\rm yr}$, so the evolution cannot be completed.
From the upper two panels, we see that the impact of $|\mathcal R|^2$ on the orbital evolution becomes more pronounced toward smaller
$p_i$ and smaller $e_i$.
However, the lower panels, tracking the accumulated orbital phases which are more directly tied to the GW phasing, reveal a more subtle dependence on the eccentricity.
The dependence on the initial semi-latus rectum remains robust: in both $\Phi_r$ and $\Phi_\phi$, the smaller $p_i$ leads to a larger
reflectivity-induced phase shift.
By contrast, the eccentricity dependence differs between the two phases.
For the radial phase $\Phi_r$, the effect of $|\mathcal R|^2$ increases with $e_i$, indicating that more eccentric orbits accumulate
a larger reflectivity-induced dephasing in the radial motion.
For the azimuthal phase $\Phi_\phi$, the trend depends on $p_i$: when $p_i< 8.6$, the smaller $e_i$ yields a larger phase shift,
whereas when $p_i\gtrsim 8.6$ the dependence reverses and the larger $e_i$ produces a larger azimuthal dephasing.
From the Fisher information perspective, the stronger waveform sensitivity
to $|\mathcal R|^2$ typically translates into tighter expected
constraints, making this figure a useful preliminary diagnostic for the parameter-estimation performance anticipated in our later MCMC run.
Therefore, we may anticipate that the smaller $p_i$ leads to tighter constraints on $|\mathcal R|^2$.
The dependence on $e_i$ is more complicated; however, judging solely from the upper two panels, the constraints may be stronger
at  a small $e_i$.

\begin{figure}[tbp]
	\centering
	\begin{minipage}[b]{0.49\textwidth}
		\centering
		\includegraphics[width=\textwidth]{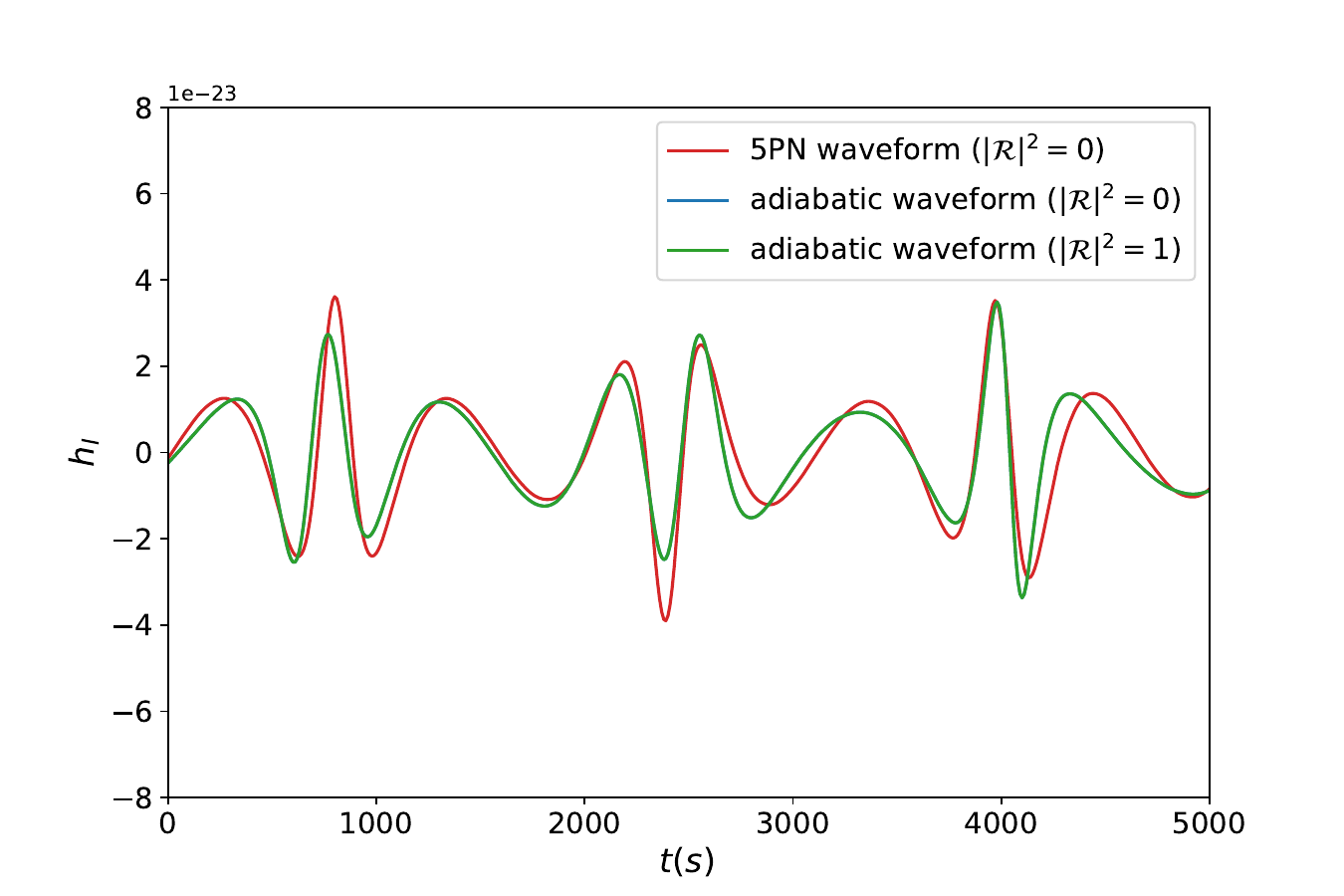}
	\end{minipage}
	\begin{minipage}[b]{0.49\textwidth}
		\centering
		\includegraphics[width=\textwidth]{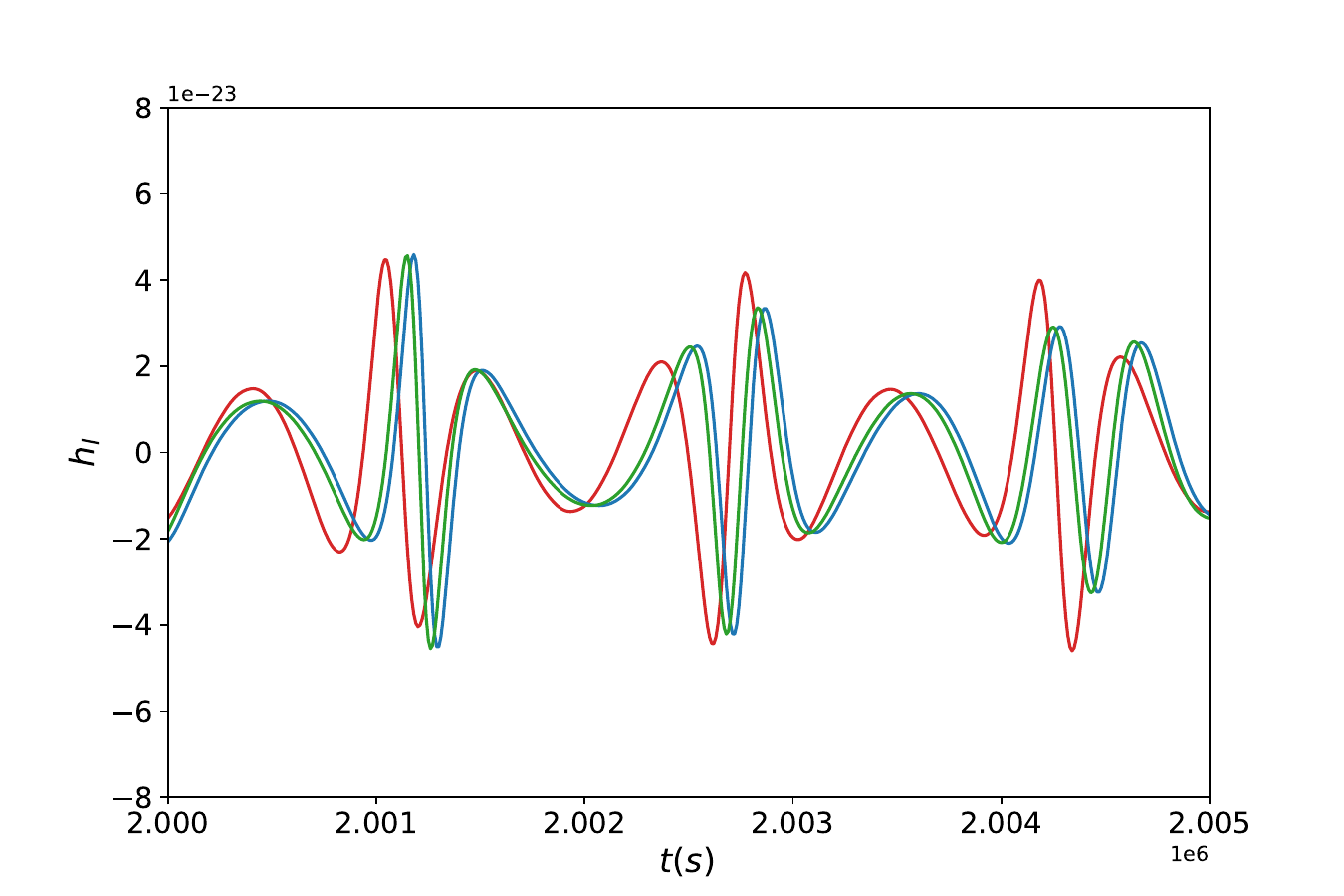}
	\end{minipage}	
	
	\caption{The comparison of the 5PN waveform (the reflectivity $|\mathcal{R}|^2=0$) with the adiabatic waveform ($|\mathcal{R}|^2=0$ and $|\mathcal{R}|^2=1$).
		Here $(M,\mu,a,p,e) = (10^{6}M_\odot, 10M_\odot, 0.95M, 13M, 0.4)$.
		The sky and viewing angles are $(q_k,\phi_k,q_s,\phi_s) = (\pi/4, \pi/3, \pi/2, \pi/3)$, the luminosity distance is $D_L = 3$ Gpc, and the initial orbital phases are $(\Phi_{\phi0},\Phi_{r0}) = (\pi/3, \pi)$. The left panel shows the initial GW waveform, and the right panel shows the waveform after $2\times10^{6} \,\mathrm{s}$.			
	}
	\label{fig:waveform}	
\end{figure}

Having established how the reflectivity modifies the orbital evolution, a natural next step is turning to its impact on the EMRI  waveform.
In Fig.~\ref{fig:waveform} we display the LISA $I$-channel response $h_I$ to the GW. Redshift effects are neglected here. As seen in the figure, at early times ($t\sim 0$) the waveform computed with the maximal reflectivity $|\mathcal{R}|^2=1$ is indistinguishable from the no tidal heating case $|\mathcal{R}|^2=0$, whereas after $2\times10^{6}\,\mathrm{s}$, a clear phase shift develops. This indicates that the waveform mismatch can be used to diagnose the presence of tidal heating corrections. Moreover, the 5PN model is inadequate for rapidly spinning backgrounds with $a=0.95M$ and sizable eccentricity $e_i=0.4$: even if its initial phase is aligned with that of the adiabatic waveform, after $2\times10^{6}\,\mathrm{s}$ it exhibits a substantial phase shift relative to the adiabatic signal and therefore fails to capture EMRI waveforms accurately. Consequently, the 5PN waveform is unsuitable for probing the tidal heating, which is most pronounced for $a \sim M$ with orbits near the separatrix.
 Here we only offer a brief qualitative remark; a full treatment of tidal heating and its impact on EMRI orbits and waveforms are beyond the scope of this work. Interested readers may consult Refs. \cite{datta2020tidal,datta2024tidal}.

\section{Bayesian inference for EMRI parameters}
\label{sec:3}
\subsection{Method}

In this subsection, we briefly introduce the setup and pipeline of our MCMC analysis.
To assess the similarity between two waveforms, it is  convenient to introduce the
noise–weighted inner product between any two waveforms $h_a(t)$ and $h_b(t)$ 
\begin{equation}
\left( h_a \mid h_b \right)
= 2 \int_{0}^{\infty} \mathrm{d}f\,
\frac{\tilde{h}_a^{*}(f)\,\tilde{h}_b(f) + \tilde{h}_a(f)\,\tilde{h}_b^{*}(f)}{S_n(f)} \, ,
\end{equation}
where the tildes denote Fourier transforms and the asterisk denotes the complex conjugation. $S_n(f)$  is  the noise power spectral density of
LISA detector~\cite{Robson:2018ifk}, which is stationary and Gaussian.
 The optimal SNR of the waveform $h_a(t)$ is then $\rho^2 = (h_a|h_a)$; in all examples below we fix the luminosity
distance to obtain the injected SNR of $\rho=50$.
The corresponding mismatch is 
\begin{equation}
	\mathcal{M}=1-\frac{\left( h_a \mid h_b \right)}{\sqrt{\left( h_a \mid h_a \right)\left( h_b \mid h_b \right)}}.
\end{equation}

With the definition of inner product, the likelihood of an observed GW signal $d$ for a
given parameter vector $\Theta$ is
\begin{equation}
	\mathcal{L}(d|\Theta)
	\propto \exp\!\left[
	-\frac{1}{2}\bigl(d-h(\Theta)\big|
	d-h(\Theta)\bigr)
	\right] ,
	\label{eq:likelihood}
\end{equation}
where the proportionality sign indicates that an overall normalization
constant  has been absorbed. The Bayes’ theorem provides the
posterior distribution of the parameters
\begin{equation}
	p(\Theta|d)
	= \frac{\mathcal{L}(d|\Theta)\,
		\pi(\Theta)}
	{\mathcal{Z}(d)} \, ,
	\label{eq:posterior}
\end{equation}
where $\pi(\Theta)$ denotes the prior,
and $\mathcal{Z}(d)$ is the Bayesian evidence.  Since we
do not perform model selection in this work, the evidence plays only
the role of an overall normalization.

\begin{table}[htbp!]
		\caption{The injected values and prior distributions on the waveform parameters 
		used for the MCMC posterior sampling.
		Here $\mathcal{U}$ denotes the uniform distribution and $\mathcal{P}$ denotes the power-law distribution.\label{tab:priors}}
	\centering
	\begin{tabular}{c|c|c|c}
		\hline
		\hline
		{Parameters} & Description&Injected values & {Priors ($\delta=0.01$)} \\ 
		\hline
		\hline
		$\ln (M/M_\odot)$ &Mass of central object &$\ln(10^6)$& $\mathcal U[\ln M * (1- \delta), \ln M * (1+ \delta)]$ \\ \hline
		$\ln (\mu /M_\odot)$ & Mass of secondary BH& $\ln(10)$& $\mathcal U[\ln \mu^* (1- \delta), \ln \mu^* (1+ \delta)]$ \\ \hline
		$a$ & Spin of central object (M)&$0.95$& $\mathcal U[a* (1- \delta), a* (1+ \delta)]$ \\ \hline
		$p_i$ & Initial semi-latus rectum (M)&$8.5$& $\mathcal U[p_i * (1- \delta), p_i * (1+ \delta)]$ \\ \hline
		$e_i$ & Initial eccentricity&$0.4$& $\mathcal U[e_i * (1- \delta), e_i * (1+ \delta)]$ \\ \hline
		$D_L$ &Distance (Gpc)&3&  $\mathcal P[0.01, 10]$ \\ \hline
		$\cos \theta_S$ &Polar sky localization angle&0&  $\mathcal U [-1, 1]$ \\ \hline
		$\phi_S$ &Azimuthal sky localization angle&$\pi$&  $\mathcal U[0, 2 \pi]$ \\ \hline
		$\cos \theta_K$ & Polar  orientation angle&$\sqrt 2/2$&  $\mathcal U[-1, 1]$ \\ \hline
		$\phi_K$ &Azimuthal orientation angle&$\pi/3$&  $\mathcal U[0, 2 \pi]$ \\ \hline
		$\Phi_{\phi_0}$ & Initial azimuthal orbital phase&$\pi/3$& $\mathcal U[0, \pi]$ \\ \hline
		$\Phi_{r_0}$ & Initial radial orbital phase&$\pi$& $\mathcal U[0, 2 \pi]$ \\ \hline
		$|\mathcal R|^2$ & Reflectivity&0& $\mathcal U[-1, 1]$ \\ 
		\hline
		\hline
	\end{tabular}
\end{table}

Following the strategy of Ref.~\cite{Strub:2025dfs}, one may iteratively repeat the parameter search and retain the configuration that maximizes the convolution power, thereby progressively narrowing the priors of the intrinsic parameters $(M,\mu,p_i,e_i,|\mathcal R|^2)$ and recentering them around their estimated values. Since the primary goal of this work is to characterize the posterior distribution of the reflectivity $|\mathcal R|^2$, we do not perform this preliminary optimization step explicitly. Instead, we assume that an equivalent narrowed prior range for these intrinsic parameters has already been obtained, and start our inference directly from these tightened priors.
Our parameters and prior settings given in Table~\ref{tab:priors} closely follow Ref.~\cite{Speri:2024qak}.
We adopt Bayesian priors tailored to an injection–recovery test. Intrinsic parameters that dominantly control the long EMRI phase evolution $( M,\mu, a, p_i,e_i)$ are assigned narrow uniform priors centered on the injected values  to accelerate the convergence without materially truncating the posterior. Extrinsic orientation angles, sky-location parameters, and initial orbital phases are assigned broad uniform priors over their natural ranges.
It should be noted here that  the reflectivity is physically bounded by $0 \le |\mathcal{R}|^2 \le 1$, but in our sampling scheme we allow $|\mathcal{R}|^2$ to take negative values and assign it a uniform prior on $[-1,1]$. After completing the sampling, we discard all samples with $|\mathcal{R}|^2 < 0$. This strategy improves the sampling efficiency but does not alter the resulting posterior distribution~\cite{Speri:2024qak}. 
The initial states of the chains are generated by drawing from a multivariate Gaussian distribution centered on the injected parameter values, with the covariance matrix taken from our initial guess. During sampling, we employ a mixture of stretch moves~\cite{2013PASP125306F} and adaptive Gaussian moves~\cite{haario2001adaptive}, each used with $50\%$ probability, to enhance the exploration efficiency of the MCMC chains. 
We implement the above multi-chain  MCMC sampling by using \textbf{Eryn}~\cite{michael_katz_2023_7705496}.

\begin{figure}[tbp]
	\centering
	\begin{minipage}[b]{0.5\textwidth}
		\centering
		\includegraphics[width=\textwidth]{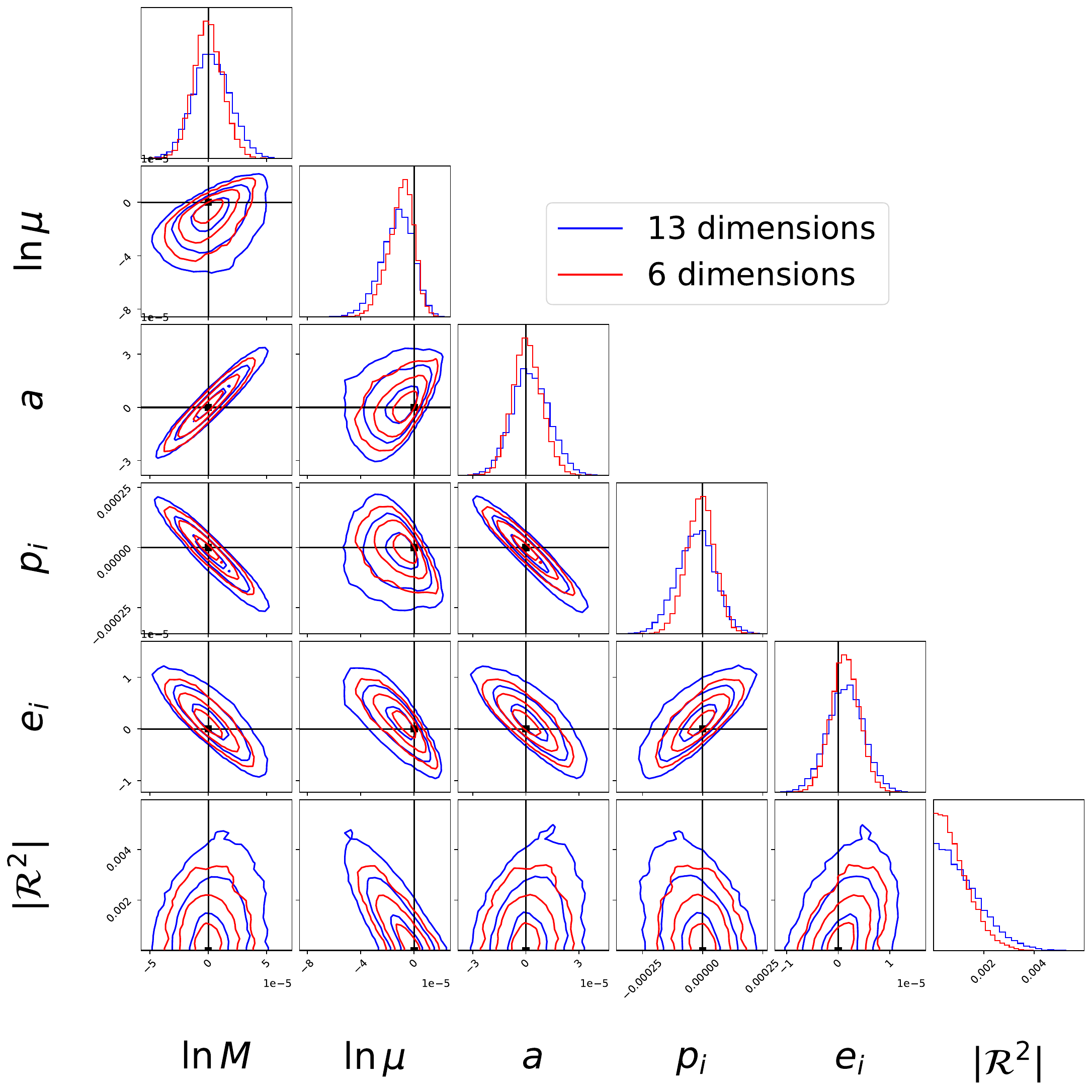}
	\end{minipage}

	\caption{
		The posterior distributions for six parameters obtained from the MCMC analysis in the full 13-dimensional parameter space (blue) and in a reduced 6-dimensional parameter space (red) with the optimal ${\rm SNR}=50$. The injected values are listed in Table~\ref{tab:priors}. For the 13-dimensional run, we sample the full 13-dimensional parameter space, but only the six parameters of interest are shown and the remaining seven parameters are omitted for clarity. For the 6-dimensional run, the remaining seven parameters are fixed to their injected values, and the sampling is performed only in the 6-dimensional subspace. The posteriors are inferred from a simulated two-year LISA observation. The vertical and horizontal lines mark the injected true value of each parameter, and the contours denote the 68.3\%, 95.4\%, and 99.7\% confidence intervals.
	}
	\label{fig:Ruduced_corner}	
\end{figure}

\subsection{Result}

We first address whether the constraint on the horizon reflectivity can be reliably assessed in the full 13-dimensional parameter space, or whether a reduced model involving only the six intrinsic parameters
$(\ln M,\ln \mu,a,p_i,e_i,\lvert \mathcal{R}\rvert^2)$ or fewer is sufficient to estimate the detectability. Fig.~\ref{fig:Ruduced_corner} shows a corner plot by comparing the marginalized posteriors obtained from the MCMC sampling in the full 13-dimensional space (blue) with those from the reduced 6-dimensional space (red).  We find that restricting to the reduced parameter set does not qualitatively alter the posterior shapes or the inferred correlations among the intrinsic parameters, but it does lead to a noticeable tightening of the posteriors by removing degeneracies associated with the marginalization over the other seven parameters. For instance, the best $68.3\%$ upper bound of reflectivity  decreases from $|{\cal R}|^2_{68.3\%}=1.49\times 10^{-3}$ in the 13-dimensional analysis to $1.11\times10^{-3}$ in the 6-dimensional analysis. While such a reduction would be warranted if the omitted parameters were independently well constrained in real observations, we currently lack strong prior knowledge of them; we therefore adopt the full 13-dimensional inference as a more conservative estimate of the constraints on $\lvert \mathcal{R}\rvert^2$.

\begin{figure}[tbp]
	\centering
	\begin{minipage}[b]{0.99\textwidth}
		\centering
		\includegraphics[width=\textwidth]{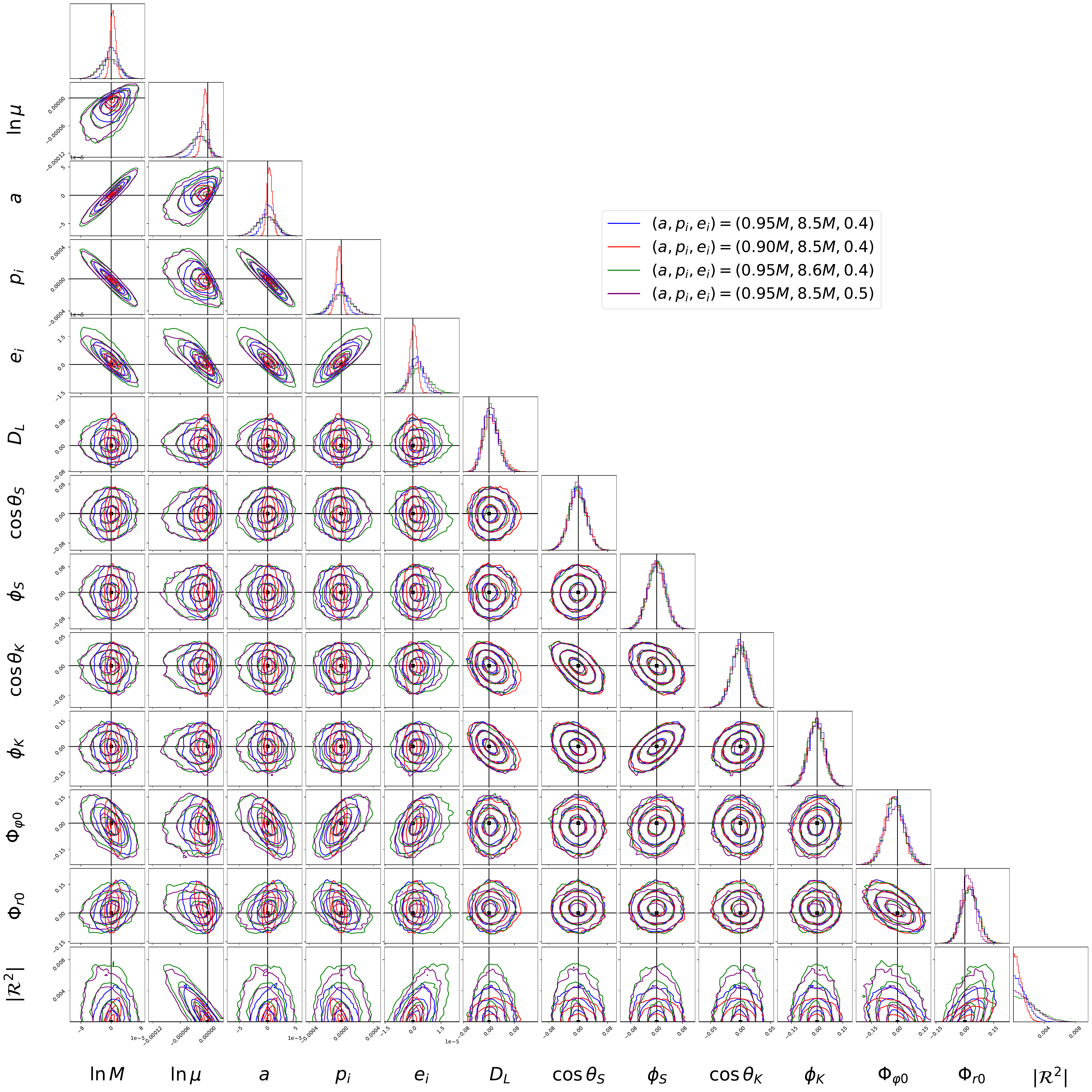}
	\end{minipage}

	\caption{		
		The posterior distributions for the EMRI source parameters obtained from the MCMC analysis with the injected value of the reflectivity $|\mathcal{R}|^{2}=0$ and optimal ${\rm SNR}=50$. The parameters varied in the inference are the MBH spin $a$, initial semi-latus rectum $p_i$ and initial eccentricity $e_i$, while the injected values of other parameters and the priors are listed in Table~\ref{tab:priors}. The posteriors are inferred from a simulated two-year LISA observation. The vertical and horizontal lines mark the injected true value of  each parameter, and the contours denote the 68.3\%, 95.4\%, and 99.7\% confidence intervals.
	}
	\label{fig:corner13D}	
\end{figure}

The posterior distributions of the 13-dimensional EMRI parameters, including the reflectivity, are summarized in Fig.~\ref{fig:corner13D} for four representative injections with different intrinsic orbital configurations $(a,p_i,e_i)$, as indicated by the colored lines in the legend. 
From this figure, we find that the EMRI parameters can be consistently recovered across all four injections: the posterior peaks lie close to the injected values, and the injected parameters are typically contained within the $1\sigma$ confidence interval. Focusing on the fiducial configuration $(a,p_i,e_i)=(0.95M,\,8.5M,\,0.4)$ (blue), the six intrinsic parameters that dominantly control the long-time phase evolution are determined with high accuracy, yielding $\ln M =13.8155131^{+1.65\times10^{-5}}_{-1.53\times 10^{-5}}$, $\ln \mu = 2.3025734^{+1.09\times10^{-5}}_{-1.32\times10^{-5}}$, $a = 0.9500025^{+1.08\times10^{-5}}_{-9.98\times10^{-6}}$, $p_i = 8.4999741^{+7.40\times10^{-5}}_{-8.01\times10^{-5}}$ and  $e_i = 0.4000016^{+3.37\times10^{-6}}_{-3.56\times10^{-6}}$, where the typical uncertainties  are of order $\mathcal{O}(10^{-5})$. By contrast, the other  seven parameters exhibit noticeably broader posteriors: $D_L = 0.6840199^{+0.0253}_{-0.0213}$,
$\cos\theta_S = -0.0006565^{+0.0223}_{-0.0232}$,
$\phi_S = 3.1432197^{+0.0260}_{-0.0278}$,
$\cos\theta_K = 0.7053163^{+0.0135}_{-0.0137}$,
$\phi_K = 1.0469281^{+0.0435}_{-0.0451}$,
$\Phi_{\phi 0} = 1.0306037^{+0.0540}_{-0.0516}$,
and $\Phi_{r 0} = 3.1672203^{+0.0435}_{-0.0425}$, in which the typical uncertainties  are of order $\mathcal{O}(0.01)$.

For the reflectivity, the marginalized posterior is strongly one-sided and accumulates near $|{\cal R}|^2=0$, indicating that the data are consistent with a non-reflective horizon in these injections. We therefore quote an upper limit rather than a symmetric error bar. For the $(a,p_i,e_i)=(0.95M,8.5M,0.4)$ case, the best $68.3\%$ upper bound is $|{\cal R}|^2_{68.3\%}=1.49\times10^{-3}$.
Parameter correlations can be seen from the marginal posterior distributions in the corner plot.
Comparing the four injections, we observe that the constraining capacity on $|{\cal R}|^2$ depends sensitively on the orbital configuration. 
For the injected values $(p_i,e_i)$, reducing the spin from $a=0.95M$ to $a=0.90M$ (red) leads to a tighter constraint: the $|{\cal R}|^2$ posterior becomes narrower  with  $|{\cal R}|^2_{68.3\%}$ changing from $|{\cal R}|^2_{68.3\%}=1.49\times 10^{-3}$ to $1.00\times 10^{-3}$.
On the contrary, increasing the initial semi-latus rectum from $p_i=8.5M$ to $8.6M$ (green)  weakens the constraint, with  the  bound moving  to $|{\cal R}|^2_{68.3\%}=2.53\times 10^{-3}$.
Finally, increasing the initial eccentricity from $e_i=0.4$ to $0.5$ (purple) 
also weakens the constraint, with the  bound moving  to $|{\cal R}|^2_{68.3\%}=2.25\times 10^{-3}$. Overall, Fig.~\ref{fig:corner13D} demonstrates that while the intrinsic EMRI parameters are robustly recovered across all injections, the constraining  capacity on the horizon reflectivity depends on the configuration and become noticeably weaker when the inferred $|{\cal R}|^2_{68.3\%}$ posterior broadens and its upper credible limit increases.

\begin{figure}[tbp]
	\centering
	\begin{minipage}[b]{0.49\textwidth}
		\centering
		\includegraphics[width=\textwidth]{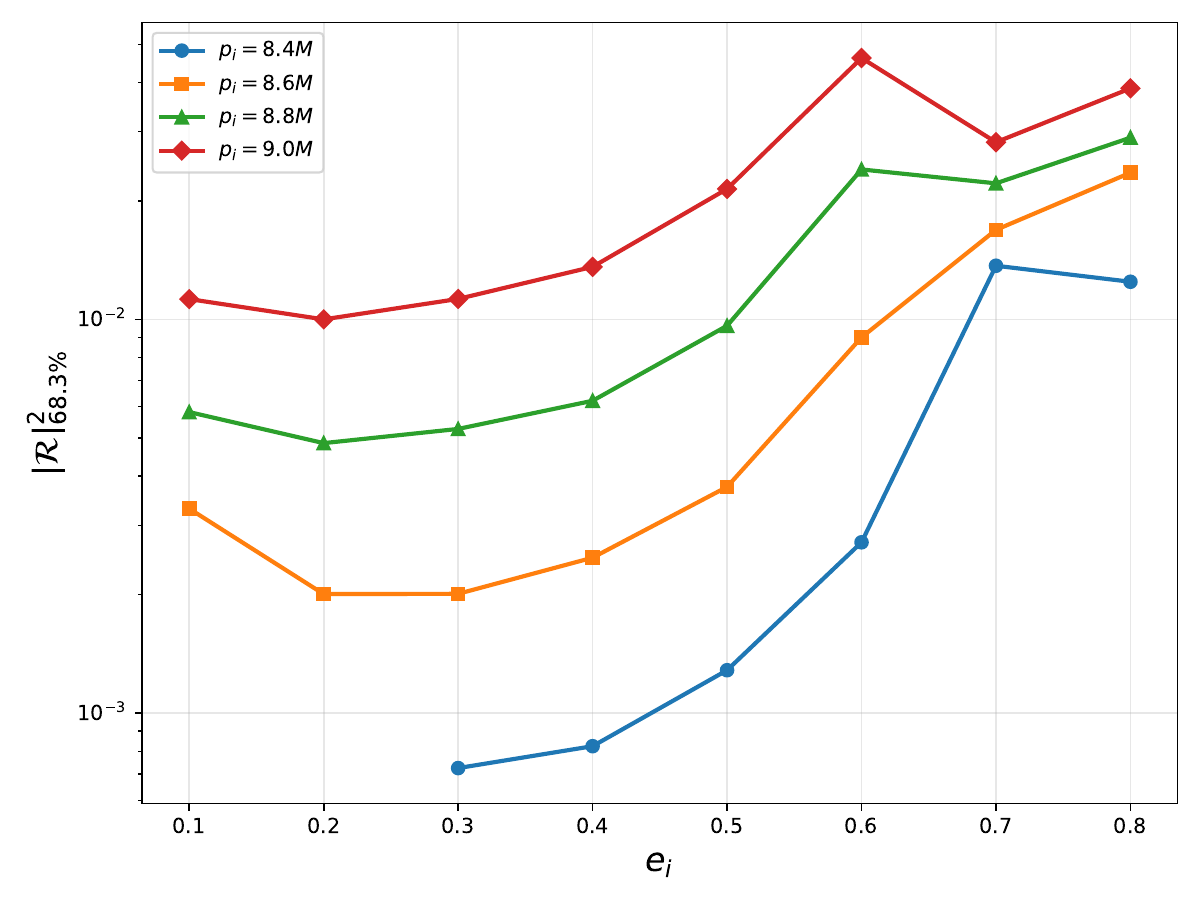}
	\end{minipage}

	\caption{The best $68.3\%$ upper bound on the reflectivity $|{\cal R}|^2$ as a function of initial eccentricity $e_{i}$ with the initial semi-latus rectum $p_{i}$, i.e., $8.4M$ (blue), $8.6M$ (orange), $8.8M$ (green) and $9.0M$ (red), respectively. The injected values of all other parameters are listed in Table~\ref{tab:priors}.
	}
	\label{fig:zhexiantu}	
\end{figure}

We now quantify more systematically how the orbital configuration controls the constraining  capacity on the horizon reflectivity, which is shown in Fig.~\ref{fig:zhexiantu}. Motivated by the configuration dependence seen in the previous section, we perform a scan over the $(p_i,e_i)$ plane by fixing the spin to $a=0.95M$. For each grid point, we summarize the best $68.3\%$ upper bound on $|{\cal R}|^2$. Two robust trends emerge. First, for a fixed eccentricity, the uncertainty grows monotonically with the increase of the initial semi-latus rectum $p_i$: the larger $p_i$ systematically yields weaker constraints (the larger $|{\cal R}|^2_{68.3\%}$). This behavior is evident across the full range of $e_i$, indicating that moving the orbit outward (hence making it less relativistic) degrades the ability to distinguish reflectivity-induced dissipative corrections. Second, the dependence on the eccentricity is non-monotonic. For each curve of $p_i$ , $|{\cal R}|^2_{68.3\%}$ is minimized around $e_i\simeq 0.2$ and increases again at higher eccentricities, so that a larger $e_i$  does not necessarily improve the constraint. Note that the $(p_i,e_i)=(8.4M,0.1)$ and $(8.4M,0.2)$ cases are absent because the corresponding inspiral time is shorter than two years and therefore does not satisfy our assumed observation duration. These findings are consistent with the qualitative expectations drawn from Fig.~\ref{fig:orbit}. There we observed that tidal heating effects strengthen as $p_i$ decreases, so it is natural that smaller $p_i$ leads to tighter reflectivity constraints. By contrast, Fig.~\ref{fig:orbit} does not uniquely determine how the constraint should vary with $e_i$. Fig.~\ref{fig:zhexiantu} indicates that moderately eccentric orbits are the most constraining: $|{\cal R}|^2_{68.3\%}$ reaches its minimum around $e_i\simeq0.2$, and a larger $e_i$ actually degrades the constraint. 
In conclusion,  we see that LISA can constrain the tidal heating at the level of $\sim 10^{-3}$--$ 10^{-4}$ with the spin of central object $a=0.95M$.

\begin{figure}[tbp]
	\centering
	\begin{minipage}[b]{0.99\textwidth}
		\centering
		\includegraphics[width=\textwidth]{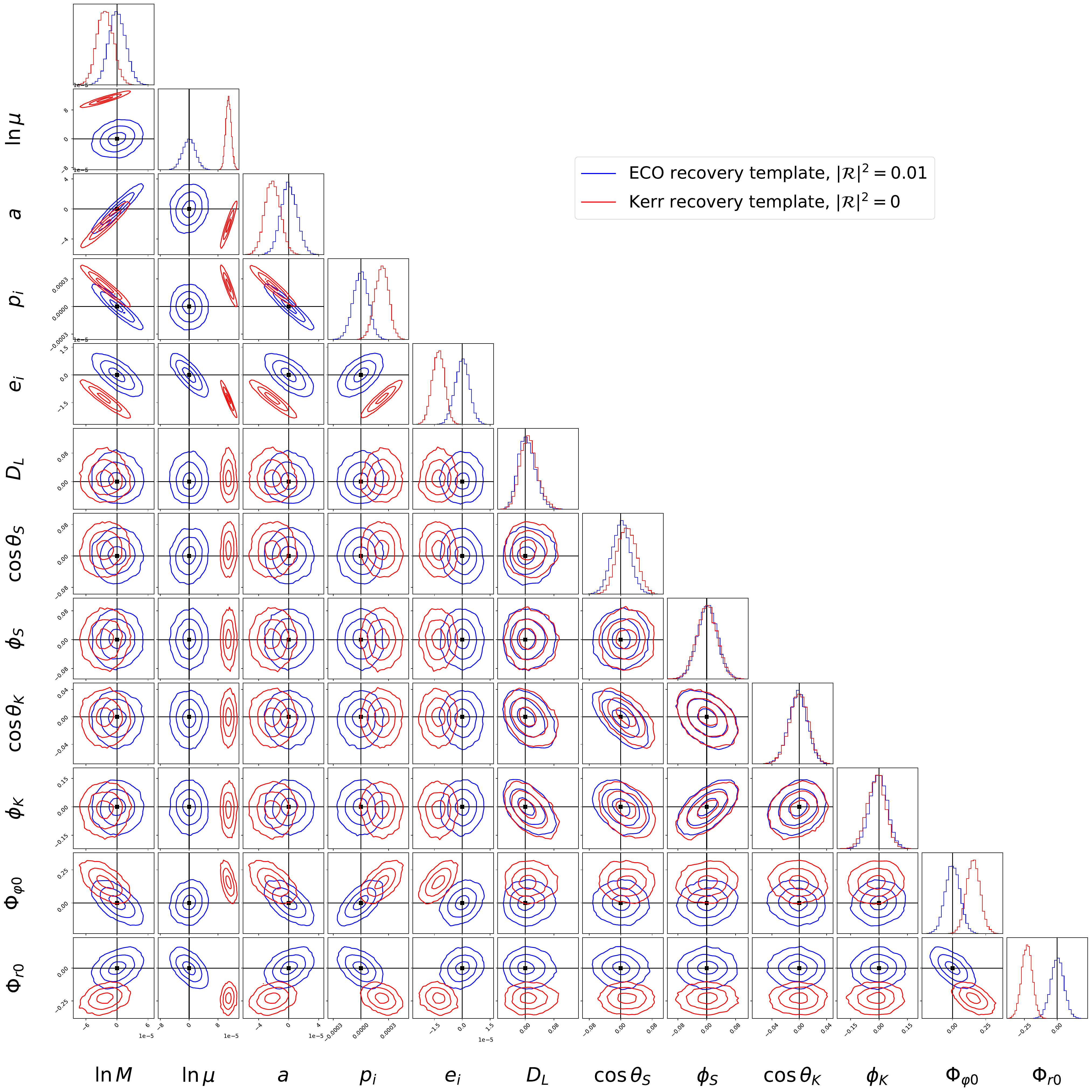}
	\end{minipage}

\caption{The posterior distribution of an EMRI system with the reflectivity $|\mathcal{R}|^2=0.01$ (blue) and $|\mathcal{R}|^2=0$ (red) for the optimal ${\rm SNR}=50$. For both cases, the injected value $|\mathcal{R}|^2$ is $0.01$, and the injected values of other parameters and the priors are listed in Table~\ref{tab:priors}. The diagonal panels show the marginalized one-dimensional posteriors, while off-diagonal panels show the two-dimensional joint posteriors. The posteriors are inferred from a simulated two-year LISA observation. The vertical and horizontal lines mark the injected true value of  each parameter, and the contours denote the 68.3\%, 95.4\%, and 99.7\% confidence intervals.}

	\label{fig:corner13DBoundary}	
\end{figure}

Finally, to quantify the level of systematic bias induced by waveform mismodeling, we perform an injection--recovery study in which the injected signal is generated with a non-zero horizon reflectivity, while the recovery is carried out with two alternative template assumptions. Specifically, we analyze the same data by using (i) a reflectivity-aware template fixed to the injected value $|\mathcal{R}|^2=0.01$ (blue), and (ii) a template that neglects the reflectivity by enforcing a perfectly absorbing horizon, $|\mathcal{R}|^2=0$ (red). The resulting posterior distributions are compared in Fig.~\ref{fig:corner13DBoundary}. We find that the recovery that ignores the reflectivity exhibits clear systematic biases in the intrinsic parameters that dominantly control the long-time EMRI phase evolution, most notably $(\ln M,\ln\mu,a,p_i,e_i)$, together with correlated shifts in the initial phases $(\Phi_{\phi 0},\Phi_{r0})$. By contrast, the extrinsic parameters (i.e., the distance, sky location and orientation angles) remain broadly consistent between the two analyses, indicating that the dominant impact of reflectivity enters through the inspiral orbits rather than through the geometrical projection.
This behavior admits a straightforward physical interpretation. In an EMRI, essentially all of the statistical constraining capability comes from the phase coherence accumulated about $10^{6}$ cycles. A non-zero reflectivity modifies the effective dissipation  by altering the horizon boundary condition, thereby changing the radiation-reaction-driven evolution of the orbits.

\section{Conclusion} 
\label{sec:4}

EMRIs are among the most promising sources for space-based GW detectors. Their long-time, highly coherent signals provide a powerful tool to probe strong-field dynamics.
Motivated by this property, in this work we investigated how EMRI
observations can constrain the horizon dissipation. 

We began by examining how a small departure from the Kerr absorbing boundary condition imprints itself on the
orbital evolution and the resulting waveform. Concretely, we compared a reference Kerr inspiral
($|\mathcal{R}|^{2}=0$) to an otherwise identical inspiral with a nonzero reflectivity, and tracked the induced
changes in the inspiral trajectories and the accumulated phases. This comparison reveals
a  trend: the inspirals with a smaller initial semi-latus rectum $p_i$ exhibit a markedly stronger imprint from the horizon-flux contribution, and hence a larger
reflectivity-induced dephasing. By contrast, the role of the eccentricity $e_i$ is more subtle. Within the range explored, changing $e_i$ induces
nontrivial and sometimes competing shifts in the accumulated orbital phases, particularly in the initial orbital phases $\Phi_{r0}$ and $\Phi_{\phi0}$, so that the effect of $e_i$ on the imprint of horizon dissipation cannot be inferred
directly from comparing the orbit alone.
Moreover, our waveform comparisons indicate that PN EMRI templates are not adequate for
detecting tidal heating effects, and that fully relativistic adiabatic waveforms are required to capture the relevant strong-field dissipation imprints reliably.

Subsequently, we carried out the Bayesian parameter estimation with a two-year LISA observation and performed
injection--recovery studies in a high-dimensional EMRI parameter space. Testing the impact of common
simplifications by comparing the full 13-dimensional inference to a reduced 6-dimensional setup with several fixed parameters, we found that fixing parameters can bias the inference and lead to overly optimistic constraints, which indicates that the fully marginalized 13-dimensional analysis is more reliable. Using this setup, we consistently recovered the Kerr limit with the marginalized posterior of $|{\cal R}|^2$ peaked
at zero, and obtained the best $68.3\%$ upper bound on the reflectivity $|{\cal R}|^2$ which depends on the intrinsic parameters.  Using two-year signals with an optimal SNR of $\rho=50$ and a fully relativistic adiabatic waveform model with the full LISA response, EMRIs can  put bounds on  $|\mathcal{R}|^2$
at the level of $10^{-3}$--$ 10^{-4}$ for a rapidly spinning central object with spin $a=0.95M$. In particular, we observed that the uncertainty increases monotonically with the increase of $p_i$ but is non-monotonic for $e_i$, as we observe that smaller semi-latus rectum and moderate eccentricity $e\simeq0.2$ yield the strongest constraints on the horizon dissipation. Furthermore, we quantified the systematic bias from waveform mismodeling via an injection--recovery test with an injected nonzero reflectivity by comparing the recovery with a nonzero-reflectivity template to a GR template without the reflectivity. The GR recovery template exhibits clear bias in the intrinsic parameters that dominantly control the secular orbit dephasing, while the extrinsic parameters remain broadly consistent. 
These results emphasize that ECO recovery  templates with the reflectivity are essential both for reliably inferring the
EMRI  parameters and for robustly probing the physical properties of the BH event
horizon.

Overall, we demonstrated that the tidal heating in eccentric EMRIs can be robustly constrained through full Bayesian inference with relativistic waveforms. This establishes EMRIs as precision probes of horizon dissipation and provides a concrete pathway for testing the tidal heating with future space-based GW observations.

\begin{acknowledgements}
We express our sincere gratitude to Zachary Nasipak for kindly providing the  flux dataset, and to  Jianjun Song for his helpful suggestions on the MCMC computations. This work made use of and references the following
software, listed in alphabetical order: Eryn~\cite{michael_katz_2023_7705496}, FastEMRIWaveforms~\cite{chapman_bird_2025_15630565}, KerrGeodesics~\cite{niels_warburton_2023_8108265}, LISAanalysistools~\cite{michael_katz_2025_17138723} and testGRwEMRIs~\cite{Speri:2024qak}. This work was supported by the National Natural Science Foundation of China (Grant Nos. 12275079, 12547143 and 12035005), the National Key Research and Development Program of China (Grant No. 2020YFC2201400) and the innovative research group of Hunan Province (Grant No. 2024JJ1006).  
This work was also supported by the financial support from Brazilian agencies Funda\c{c}\~ao de Amparo \`a Pesquisa do Estado de S\~ao Paulo (FAPESP), Funda\c{c}\~ao de Amparo \`a Pesquisa do Estado do Rio de Janeiro (FAPERJ), Conselho Nacional de Desenvolvimento Cient\'{\i}fico e Tecnol\'ogico (CNPq), and Coordena\c{c}\~ao de Aperfei\c{c}oamento de Pessoal de N\'ivel Superior (CAPES).

\end{acknowledgements}

\bibliographystyle{apsrev4-2}
\bibliography{reference}

\end{document}